\documentclass{emulateapj}
\usepackage{natbib}
\usepackage{graphicx}
\input{epsf.sty}

\slugcomment{Accepted by Astrophysical Journal}

\begin{document}
\title{THE PHYSICS OF PROTOPLANETESIMAL DUST AGGLOMERATES. VII. THE LOW-VELOCITY COLLISION BEHAVIOR OF LARGE DUST AGGLOMERATES}

\shorttitle{PROTOPLANETESIMAL DUST AGGLOMERATES VII}
\author{\scshape Rainer Schr\"apler and J\"urgen Blum}
\affil{Institut f\"ur Geophysik und extraterrestrische Physik, University of Braunschweig \\
Mendelssohnstr. 3, D-38106 Braunschweig, Germany}
\email{r.schraepler@tu-bs.de}
\and
\author{\scshape Alexander Seizinger and Wilhelm Kley}
\affil{Institut f\"ur Astronomie und Astrophysik, University of T\"ubingen \\
Auf der Morgenstelle 10, D-72076  T\"ubingen, Germany}

\shortauthors{Schr\"apler, Blum, Seizinger \& Kley}


\begin{abstract}
We performed micro-gravity collision experiments in our laboratory drop-tower using 5-cm-sized dust agglomerates with volume filling factors of 0.3 and 0.4, respectively. This work is an extension of our previous experiments reported in \citet{Beitzetal2011} to aggregates of more than one order of magnitude higher masses. The dust aggregates consisted of micrometer-sized silica particles and were macroscopically homogeneous. We measured the coefficient of restitution for collision velocities ranging from $\rm 1~cm~s^{-1}$ to $\rm 0.5~m~s^{-1}$, and determined the fragmentation velocity. For low velocities, the coefficient of restitution decreases with increasing impact velocity, in contrast to findings by \citet{Beitzetal2011}. At higher velocities, the value of the coefficient of restitution becomes constant, before the aggregates break at the onset of fragmentation. We interpret the qualitative change in the coefficient of restitution as the transition from a solid-body-dominated to a granular-medium-dominated behavior. We complement our experiments by molecular dynamics simulations of porous aggregates and obtain a reasonable match to the experimental data. We discuss the importance of our experiments for protoplanetary disks, debris disks, and planetary rings. The work is  an extensional study to previous work of our group and gives a new insight in the velocity dependency of the coefficient of restitution due to improved measurements, better statistics and a theoretical approach.
\end{abstract}
\keywords{coefficient of restitution --- fragmentation --- dust ---planetary rings ---debris disks --- impact --- solar system: formation}

\maketitle

\section{INTRODUCTION} \label{kap:INTRO}
There is observational evidence that cm-sized particles exist in protoplanetary disks (PPDs) \citep[see e.g.][]{Wilneretal2005}. A recent model of the protoplanetary-dust growth, based on laboratory experiments, has shown that cm-sized particles can be formed by direct collisional sticking \citep{Guettleretal2010,Zsometal2010}. \citet{Guettleretal2010} also found that collisional compaction can lead to filling factors of up to 0.57. Compaction in fragmenting collisions with mass transfer was also found by \citet{Kotheetal2010}, who confirmed the model by \citet{Guettleretal2010}.

If there were even larger solid particles available in PPDs, then the growth could commence through a fragmentation-coagulation cycle, leading to dust-aggregates sizes in the planetesimal size range \citep{Windmarketal2012}. As the dust growth in PPDs starts with (sub-)$\rm \mu m$ dust grains, it is natural to assume that the resulting macroscopic bodies are agglomerates of the microscopically small dust grains. Many bodies in debris disks and planetary rings are also expected to consist of such granular material. Therefore, it is interesting and important to know the collision behavior of very large dust aggregates. A first approach to the $>$cm size range was established by \citet{Beitzetal2011}, who investigated collisions among 2-cm-sized dust aggregates. Here, we present follow-up experiments with 5-cm-sized dust aggregates, which are more than an order of magnitude larger in mass.

In Section \ref{kap:EXACT} and \ref{sec:expres}, we describe our experimental approach and the experimental results. Section \ref{sec:model} explains the numerical model to understand the physics in dust-aggregate collisions, and Section \ref{sec:applications} gives some astrophysical applications for low-velocity collisions of granular bodies. Finally, Section \ref{kap:COCON} summarizes our results.

\section{EXPERIMENTAL APPROACH} \label{kap:EXACT}

In this Section, we describe the experimental methods applied for the determination of the coefficient of restitution and the fragmentation threshold of 5-cm-sized dust agglomerates.

\subsection {Preparation of the Dust Agglomerates}
The dust material for the production of large aggregates was pure SiO$_2$ powder, consisting of 0.1-10$\mu$m sized irregular grains. Owing to the fact that direct growth can commence through the mm-size range, upon which new growth process (e.g., fragmentation with mass transfer) occur, we expect that large dust agglomerates in PPDs possess a hierarchic structure and are agglomerates of agglomerates. Therefore, prior to the compression into large dust aggregates, the dust powder was being sieved through a mesh with a width of 0.5 mm to avoid larger aggregates. As shown by \citet{Weidling2012}, the sieving process produces dust agglomerates with a filling factor of 0.35.
The sieved dust was then filled into a hollow steel cylinder with 5 cm diameter and was then slowly compressed with a brass piston. The volume filling factor of the compressed dust aggregates was adjusted by filling the cylinder with a defined dust mass $m$ and by compressing the sample to a pre-determined height of 5 cm. After the compression, the bottom of the steel cylinder was removed and the dust agglomerate was pushed out of the cylinder using the piston.

\subsection{Morphology of the Dust Agglomerates}\label{sect:XRT}
\begin{figure}[!thb]
    \center
      \includegraphics[width=0.4\textwidth]{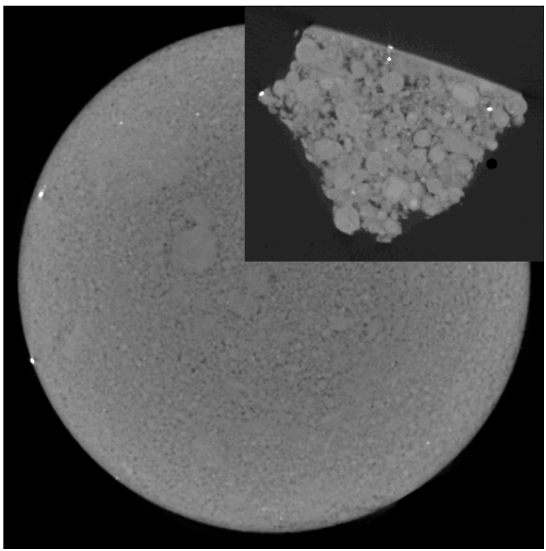}
    \caption{\label{fig:Slice} Reconstruction of an XRT image showing a radial slice perpendicular to the cylinder axis in the center of the dust agglomerate (window size: 5.9 cm $\times$ 5.9 cm). Inset: cut-out of a piece of the dust agglomerate, including the cylinder mantle (window size: 0.63 cm $\times$ 0.54 cm).}
\end{figure}

To examine the inner structure of our dust agglomerates, we performed x-ray tomography (XRT) measurements.  Figure \ref{fig:Slice} shows a plane perpendicular to the cylinder axis of 5 cm diameter and shows granularity, which represents the original sieved dust aggregates. This sub-structure is not visible in some areas of a few mm in radius. We think that this is caused by slightly  inhomogeneous compression. To increase the resolution of the XRT, we cut a  mm-sized piece out of a larger dust aggregate, which included the cylinder mantle, and repeated the XRT measurements with a voxel size of 17$\mu$m $\times$ 17$\mu$m $\times$ 17$\mu$m. In the inset of Figure \ref{fig:Slice}, a cut through the center of the sample is displayed, with the former cylinder mantle at the top of the inset picture. On the  cylinder mantle, the agglomerate-of-agglomerates structure is not present. The material has obviously formed a slightly densified mantle of about 85$\mu$m thickness with a filling factor of 0.35. We do not think that this slight density increase by about 10 percent has a considerable effect on the impact behavior of the dust agglomerates.

\begin{figure}[!thb]
   \center
    \includegraphics[width=0.4\textwidth]{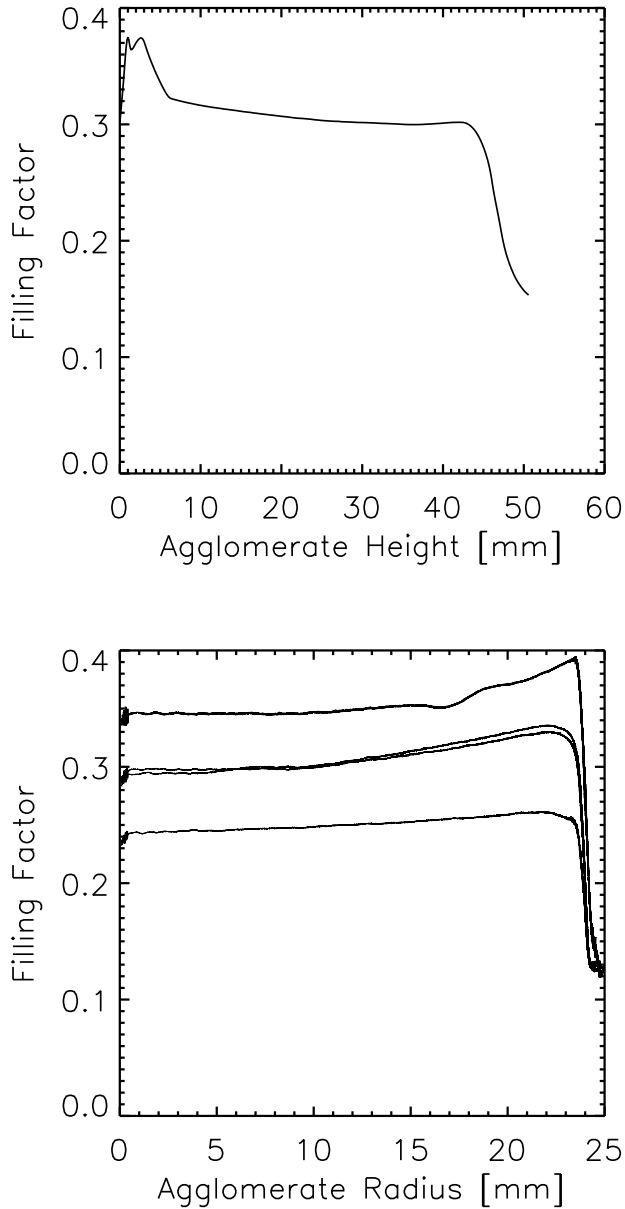} 
    \caption{\label{fig:DensPlot} Volume filling factors derived from the XRT measurements of a dust agglomerate. Top graph: the mean density of planes  perpendicular to the cylinder axis as a function of the height of the dust agglomerate. Bottom graph: the radial density distribution averaged over the azimuth and 20\% of the dust-cylinder height for five vertical positions of the dust agglomerate. The uppermost curve belongs to the top part of the dust aggregate, the three center curves (partly overlapping) belong to its central parts and the lower curve represents the lower portion of the dust agglomerate. }
\end{figure}

Figure \ref{fig:DensPlot} shows the density distribution inside the dust agglomerate. In the top graph, the mean density of planes perpendicular to the cylinder axis is shown over the full height of the agglomerate. The median volume filling factor of the dust agglomerate is $\sim 0.3$. At the top 6 mm of the dust-agglomerate cylinder, the mean density is increased to a filling factor of up to 0.38. This is the place where the piston has pushed against the agglomerate during compression. In the lower 5mm of the dust aggregate, the density of the agglomerate rapidly decreases to a volume filling factor of about 0.15. The lower graph of Figure \ref{fig:DensPlot} shows the radial density distribution averaged over the azimuth and over 20\% of the cylinder height for each curve. The upper curve was averaged over the upper 20\% of the dust agglomerate. The three curves in the center were averaged over the central 20\%-40\%, 40-60\% and 60\%-80\% slices of the dust agglomerate. Mind that these three curves have basically the same values and overlap in Figure \ref{fig:DensPlot}. The lower curve was averaged over the lower 20\% of the dust-agglomerates height. The lower graph of Figure \ref{fig:DensPlot} shows that the density of the central part of the dust agglomerate is slightly increasing from the axis to the mantle by about 5\% in volume filling factor. Here, the top and bottom curves are of no particular importance, because the dust agglomerates collide close to their midplanes. Within the accuracy of  the XRT measurements ($\sim 10-100 \mu$m), only slightly densified material on the dust-agglomerate's cylinder mantle is found, which we believe does not dramatically influence the collisional outcome.

\subsection {Collision Experiments}
\begin{figure}[!thb]
    \center
    \includegraphics[width=0.4\textwidth]{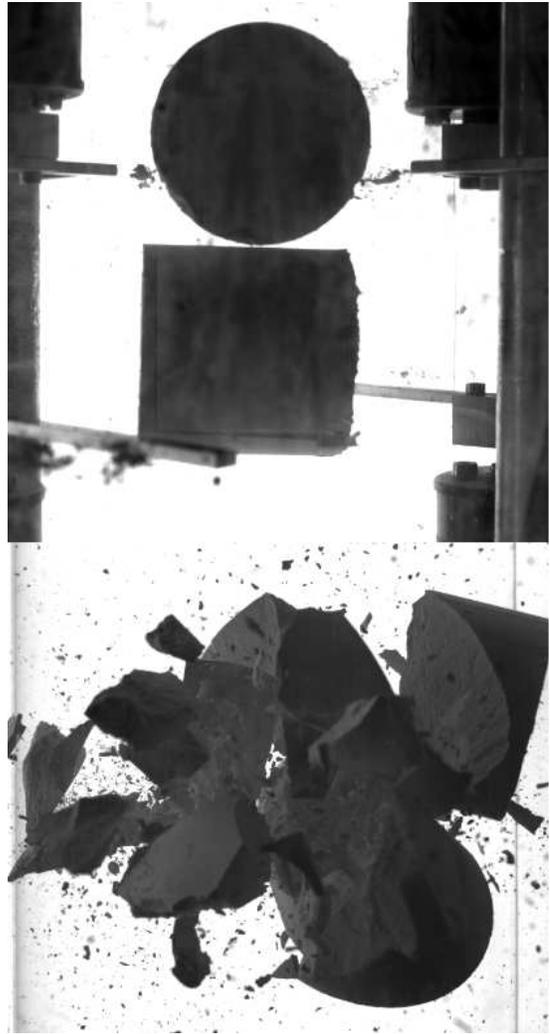} 
    \caption{\label{fig:release} Top: the two dust agglomerates shortly after release and before the collision. Both release mechanism are fully open. Bottom: the two dust aggregates fragment shortly after impact.}
\end{figure}

The collision experiments using the dust aggregates described above are performed in our 1.5-m laboratory drop tower described in \citet {Beitzetal2011}. The agglomerates were placed above each other with their symmetry axes rotated by 90 degrees (see Figure \ref{fig:release} top). Each dust agglomerate is supported by two brackets, which can be rapidly pulled away by solenoid magnets. The upper dust agglomerate is released slightly earlier than the lower one (see Figure \ref{fig:release}). The time difference $\Delta t$ between the release of the upper and lower dust agglomerate results in a relative velocity $v = g  \Delta t$ between the two bodies, where $g = 9.8 ~\rm m~s^{-2}$ is the surface acceleration of the Earth. As the drop tower is evacuated to a residual gas pressure of 100 Pa, air drag can be neglected during the $\sim$ 0.5 s free-fall time.

The resulting impacts are observed by two cameras (one high-speed camera, one camera with a normal frame rate) with 90 degrees angular distance. The cameras, which are outside the drop tower, are released such that they fall in the center-of-mass frame of the two dust aggregates.

The release mechanisms of the two dust agglomerates cause a slight rotation of the two bodies. Furthermore, the collisions were not always perfectly central, which also causes rotation. To account for these effects, we calculated the coefficient of restitution from the ratio of the square root of the total kinetic energies of the two dust agglomerates in their center-of-mass frame before and after the collision,
\begin{eqnarray}
\label{eq:cor}
e=\sqrt{\frac{ m^* v_2^2 + I (\omega_{21}^2 + \omega_{22}^2)}{{ m^* v_1^2 + I (\omega_{11}^2 + \omega_{12}^2)}}}.
\end{eqnarray}
Here, $v_1$ and $v_2$ are the relative translational velocities of the two dust aggregates before (index 1) and after (index 2) the collision, $\omega_{11}$, $\omega_{12}$, $\omega_{21}$, and $\omega_{22}$ are the mutually perpendicular angular velocity components of the colliding cylinders (the first index refers to before and after the collision, the second index describes, which component of the angular velocity is meant), and $m^* = m/2$ and $I$ are the reduced mass and the moment of inertia of the two dust aggregates. Please note, that we only consider rotation of the dust cylinders around axes perpendicular to their symmetry axis. A rolling motion of the dust aggregates, which is both difficult to observe and possesses a much smaller moment of inertia, is neglected here. The effective relative velocity, including rotation effects, of the agglomerates prior to the collision is then given by
\begin{eqnarray}
\label{eq:veff}
v_{\rm eff}=\sqrt{\frac{ m^* v_1^2 + I (\omega_{11}^2 + \omega_{12}^2)}{m^*}}.
\end{eqnarray}
This rotation of the dust aggregates is only important at the very lowest impact velocities; at higher relative velocities, the rotational motion prior to the impact is negligible with respect to the rotation after the impact.

At velocities lower than $\sim \rm 10~cm~s^{-1}$, the statistical scattering of the single measurement was very high. This was due to the fact that the free-fall time of the dust aggregates was limited to $\sim$0.5 s, which allows for a 1 $\rm cm~s^{-1}$ impact velocity a maximum distance between the dust agglomerates prior and after the impact of only 2.5 mm. This is both from the preparatory point of view and for data-analysis reasons a limiting value so that lower impact velocities were not achievable. Therefore, more experiments have been performed at the lower velocities to get a reasonable mean value for the coefficient of restitution.

\section{EXPERIMENTAL RESULTS} \label{sec:expres}
In this Section, we will present our data on the low-velocity coefficients of restitution as well as on the fragmentation velocities of dust aggregates of two different volume filling factors.

\subsection{Coefficient of Restitution}\label{kap:CofRest}
\begin{figure}[!thb]
    \center
    \includegraphics[width=0.35\textwidth]{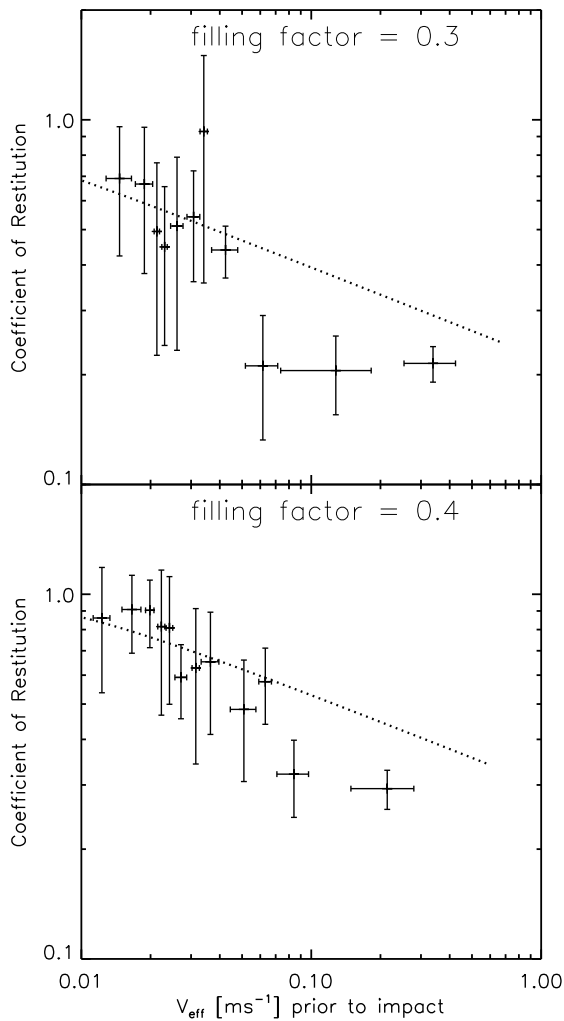}
    \caption{\label{fig:kofref}The coefficient of restitution as a function of the effective impact velocity for 5-cm sized dust agglomerates of 30\% volume filling factor (upper graph) and 40\% volume filling factor (lower graph), respectively. Each data point is the mean value of four measurements; the error bars denote the 2-$\sigma$ uncertainties of the mean values. The dotted line shows a power law with a slope of -1/4, following the solid-state model by \citet{ThNi} (see Section \ref{sec:consim}).}
\end{figure}

In Figure \ref{fig:kofref}, the coefficient of restitution of our 5-cm sized cylindrical agglomerates, according to the definition in Eq. \ref{eq:cor}, is plotted as a function of the effective impact velocity, according to Eq. \ref{eq:veff}, for volume filling factors of $\phi = 0.3$ and $\phi = 0.4$, respectively. Each data point is the mean of four measurements; the error bars denote the 2-$\sigma$ error of the mean value.

We can clearly distinguish different velocity regions: (1) At the very lowest velocities ($v_{\rm eff} \lesssim \rm 4~cm~s^{-1}$), the coefficient of restitution follows the solid-state model by \citet{ThNi} within the measurement uncertainties as shown by the error bars in Figure \ref{fig:kofref} (see Section \ref{sec:consim}). (2) At higher velocities ($\rm 4~cm~s^{-1} \lesssim v_{\rm eff} \lesssim 6~cm~s^{-1}$) for a filling factor of 0.3 and  $\rm 4~cm~s^{-1} \lesssim v_{\rm eff} \lesssim 10~cm~s^{-1}$ for a filling factor of 0.4), the coefficient of restitution decreases steeply to values of $e = 0.2$ for $\phi = 0.3$ and $e = 0.3$ for $\phi = 0.4$, respectively. (4) At even higher velocities, we observe fragmentation (see Section \ref{frag}).

\begin{figure}[!thb]
    \center
    \includegraphics[width=0.35\textwidth]{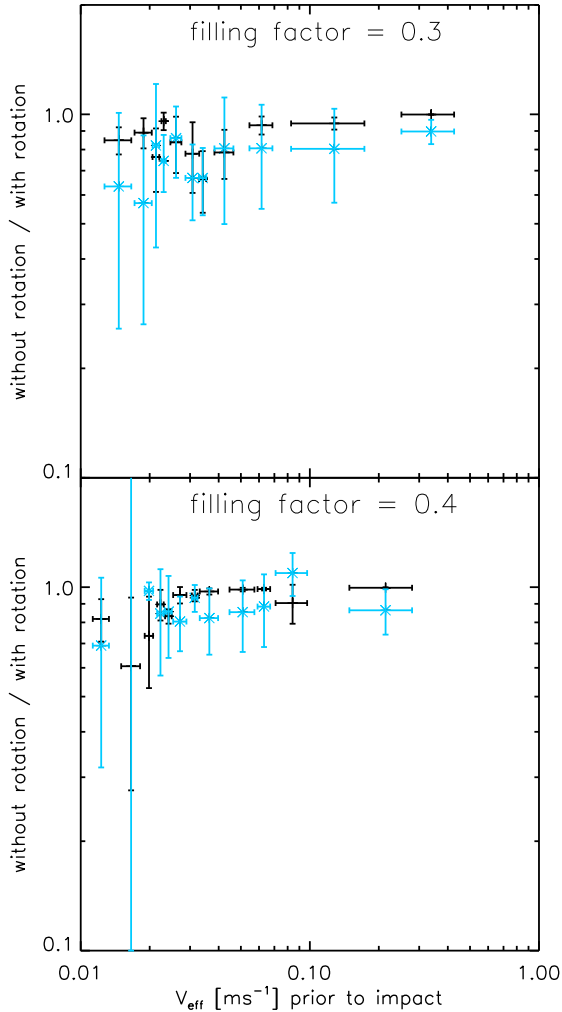}
    \caption{\label{fig:RoCont} The ratio of the effective reduced velocity (including rotation) to linear velocity prior to impact ($+$) and the ratio or the coefficient of restitution including rotational effects to the coefficient of restitution with neglected rotation ($*$).}
\end{figure}

The coefficient of restitution was measured by considering translational and rotational velocity changes. Because translational and rotational velocities are transformed into one another, depending on the impact parameter, it is not possible to split the coefficient of restitution into components of rotation and translation. To show the contribution of rotation in our measurements, we display in Figure \ref{fig:RoCont} the ratio of the translational velocity to the effective velocity prior to the impact as well as the ratio of the coefficients of restitution with and without the consideration of rotation over the effective impact velocity. We use the same averaging process as in fig. \ref{sec:consim}. Figure \ref{fig:RoCont} shows that the rotational contribution in our measurements is below 40\% over all velocities and that the influence of rotation on the coefficient of restitution increases with decreasing impact velocity. However, the data also show that the different regimes of the coefficient of restitution (see above) are not caused by rotation effects.

In contrast to our result, \citet{Beitzetal2011} found no obvious correlation between the coefficient of restitution and impact velocity. Unlike in our experiments, they used smaller spherical dust agglomerates made from monodisperse spherical grains. We used cylindrical agglomerates, because we were not able to produce homogeneous crack-free spherical dust agglomerates. Our polydisperse irregular $\rm SiO_2$ grains possess smaller contact forces between the monomer particles than the monodisperse spherical grains of \citet{Beitzetal2011}. This and the smaller filling factor in our experiments ease collisional compression and reduce the effect of randomly jammed packings during collision \citep[see e.g.][]{TorStill}. Jammed packings are stronger at higher compression velocities and increase the rigidity of an agglomerate and therewith decrease the contact area of the colliding agglomerates. Due to the model by \citet{ThNi}, this should increase the coefficient of restitution. Therefore, it is possible that jamming occurred in the experiments by  \citet{Beitzetal2011}, due to their higher filling factor and monodisperse spherical grains, which is the more pronounced at higher the collisional velocities. 

\subsection{\label{frag}Fragmentation Velocity}
\begin{figure}[!thb]
    \center
    \includegraphics[width=0.35\textwidth]{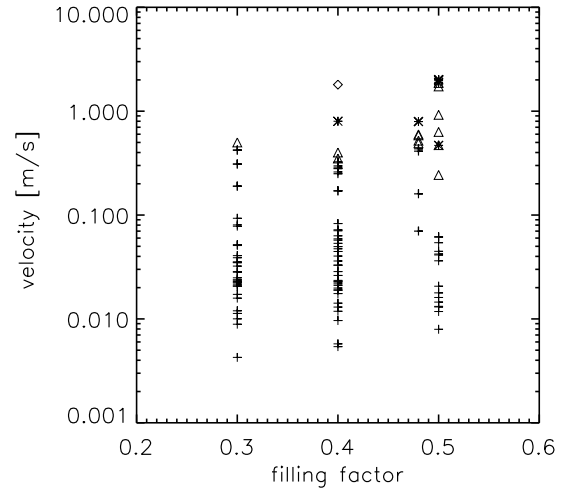}
    \caption{\label{fig:fragvel}The possible outcomes in collisions among 5-cm sized dust aggregates for different volume filling factors between $\phi = 0.3$ and $\phi = 0.48$. A $+$ denotes bouncing, $\Delta$ stands for the fragmentation of one of the two dust aggregates, $*$ and $\diamond$ describe the fragmentation of both dusty bodies, the latter with a maximum fragment mass smaller than half of the projectile mass. For comparison, the data from \citet{Beitzetal2011} for 2-cm sized spherical dust aggregates with $\phi = 0.5$, using spherical $\rm 1.5\mu m$ sized $\rm SiO_2$ monomers, are shown additionally.}
\end{figure}

Above a certain impact velocity, one or both of the colliding dust aggregates fragment so that the determination of the coefficient of restitution becomes meaningless. We measured the outcome of collisions among the 5-cm sized dust aggregates with volume filling factors of $\phi = 0.3$, $\phi = 0.4$ and $\phi = 0.48$ for velocities up to $200$~cm~s$^{-1}$ and plotted the collision results in Figure \ref{fig:fragvel}. The $+$ signs denote bouncing, $\Delta$ stands for the fragmentation of one of the two dust aggregates, whereas $*$ and $\diamond$ describe the fragmentation of both dusty bodies (in the latter case with a maximum fragment mass smaller than half of the projectile mass). For comparison, we added the data from \citet{Beitzetal2011} for 2-cm sized spherical dust aggregates with $\phi = 0.5$, using spherical $\rm 1.5\mu$m sized $\rm SiO_2$ monomers. Within the scatter of data, the threshold between bouncing and single-aggregate fragmentation seems to be independent of the volume filling factor at $v_{\rm frag} = 40 \pm 10 ~\rm cm~s^{-1}$. This is about a factor of two higher than the fragmentation threshold found by \citet{Beitzetal2011} and might be due to the different geometry (spheres vs. cylinders), the different monomer morphologies and sizes (irregular polydisperse vs. spherical monodisperse), dust-aggregate size (5 cm vs. 2 cm) or volume filling factor ($\phi \le 0.48$ vs. $\phi =0.5$).

\section{MODELING DUST-AGGREGATE COLLISIONS} \label{sec:model}

\subsection{\label{sec:consim}Continuum Theories}
Assuming that at low velocities dust agglomerates behave like solid bodies, we can apply the theory of \citet{ThNi} for the derivation of the impact-velocity dependence of the coefficient of restitution. At larger velocities, we assume that the granular material gets more and more mobilized and begins to fluidize so that a solid-state theory is no longer applicable. The model by \citet{ThNi} relies on two threshold velocities, the transition velocity between sticking and bouncing and the transition velocity between elastic and plastic material effects, $v_y$, respectively. The data shown in Figure \ref{fig:kofref} suggest that our collision velocities are clearly above both thresholds, because the coefficient of restitution is in the regime in which energy-dissipating effects dominate. In this velocity regime, \citet{ThNi} predict a decrease of the coefficient of restitution with increasing velocity and asymptotically a power-law behavior of the coefficient of restitution of the form $e \propto v^{-1/4}$. The dotted line in Figure \ref{fig:kofref} shows such a dependence, with $v_y = 0.12 \rm ~cm~s^{-1}$ for $\phi =0.3$ and $v_y = 0.4 \rm ~cm~s^{-1}$ for $\phi =0.4$, respectively. As can be seen in the comparison between the model and our data, our 5-cm sized dust aggregates can be reasonably described by the \citet{ThNi} continuum theory for effective impact velocities up to $\sim 4~\rm cm~s^{-1}$. However, for larger impact speeds, the coefficient of restitution drops below the curve predicted by \citet{ThNi} and is rather velocity independent for velocities between $\sim 10~\rm cm~s^{-1}$ and the fragmentation limit (see Figure \ref{fig:kofref}).

\subsection{\label{sec:mdsim}Molecular Dynamics Simulations of Dust-Aggregate Collisions}

To understand these different collision behaviors of dust aggregates, we use a molecular dynamics approach featuring detailed contact mechanics of microscopic silicate grains. The corresponding interaction laws have been proposed by \citet{1971RSPSA.324..301J} and \citet{1995PMagA..72..783D,1996PMagA..73.1279D}. An overview is given by \citet{DominikTielens1997} who applied the model to simulate dust agglomerate collisions for the first time in the context of planet formation. Later, \citet{2007ApJ...661..320W} presented a different approach where the same interaction laws were derived from potentials. Recently, \citet{Seizingeretal2012} proposed simple modifications to the model to better reproduce the compression behavior measured in laboratory experiments by \citet{2009ApJ...701..130G}. 

In this work the modified model of \citet{Seizingeretal2012} is used. The material parameters used here are given in Table \ref{tab:material_parameters}. They are identical to those of \citet{Seizingeretal2012} but differ from the ones used for similar simulations performed by \citet{2011ApJ...737...36W}.

\begin{table}
 \caption[]{Material Parameters of the individual monomers used in the molecular dynamics simulations.}
 \label{tab:material_parameters}
 \centering
 \renewcommand\arraystretch{1.2}
 \begin{tabular}{ll}
   \hline
   \noalign{\smallskip}
   Physical property & Silicate\\
   \noalign{\smallskip}
   \hline
   \noalign{\smallskip}
   Particle Radius $r$ (in $\mathrm{\mu m}$) & $0.6$\\
   Density $\rho$ (in g\, cm$^{-3}$) & $2.65$\\
   Surface Energy $\gamma$ (in mJ\, m$^{-2}$) & $20$\\
   Young's Modulus $E$ (in GPa)         & $54$\\
   Poisson Number $\nu$               & $0.17$\\
   Critical Rolling Length $\xi_\mathrm{crit}$ (in nm) & $2$\\
   \noalign{\smallskip}
   \hline
 \end{tabular}
\end{table}

\subsubsection{Sample Generation}
Here, we study the collisions of spherical aggregates. The samples were generated using the following procedure: we began with the regular lattice of the hexagonal closest packing (hcp) for which the volume filling factor is given by $\phi = \pi / 2\sqrt{3} \approx 0.74$. To achieve lower filling factors, we removed the proper amount of randomly selected monomers. Especially for lower filling factors, some monomers may end up disconnected from the rest of the aggregate. In the last step, we therefore removed all isolated monomers which were not connected to the main aggregate (typically less than $10^{-3}$ of the total number of monomers). Except for the removal of isolated monomers, this preparation method has been used by \citet{2011ApJ...737...36W} before.

\begin{figure}
\resizebox{\hsize}{!}{\includegraphics{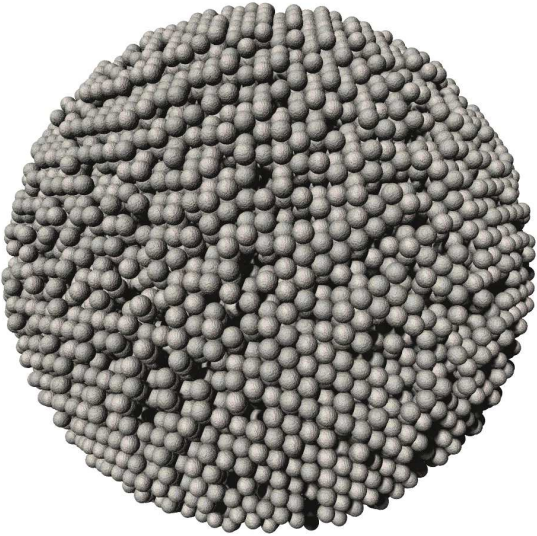}}
\caption{An example of an aggregate used for the collision simulations. It possesses a mean filling factor of $\phi = 0.55$ and consists of $\approx 40,000$ monomers.}
\label{fig:sample_aggregate}
\end{figure}

The samples we used in this work consist of approximately $4\times10^4$ monomers, with a coordination number of $\approx 8.6$, a diameter of $50\,\mathrm{\mu m}$ and a volume filling factor of $ \phi \approx 0.55$. An image of such a sample is shown in Figure \ref{fig:sample_aggregate}.

\subsubsection{Numerical Collision Experiments}
Three samples generated by the procedure described above were rotated randomly and collided head-on with impact velocities between $0.01$ and $0.5\,\mathrm{m~s^{-1}}$. In case of a bouncing event, we determined the relative velocity between the collision partners by averaging over the individual velocities of the monomers that each aggregate was composed of.

Due to the lattice structure of the samples, their relative orientation has a significant influence on the outcome of the collision. Depending on this orientation, both sticking and bouncing may occur for a given velocity. As we intended to study the dependency of the coefficient of restitution on the collision velocity, we kept the same orientation for a whole collision sequence. For each sample, eight randomly chosen orientations were examined. Thus, 24 collisions sequences have been simulated in total. Typically, not all collisions of one sequence resulted in bouncing.

\subsubsection{Results}
\begin{figure}[!h]
\resizebox{\hsize}{!}{\includegraphics{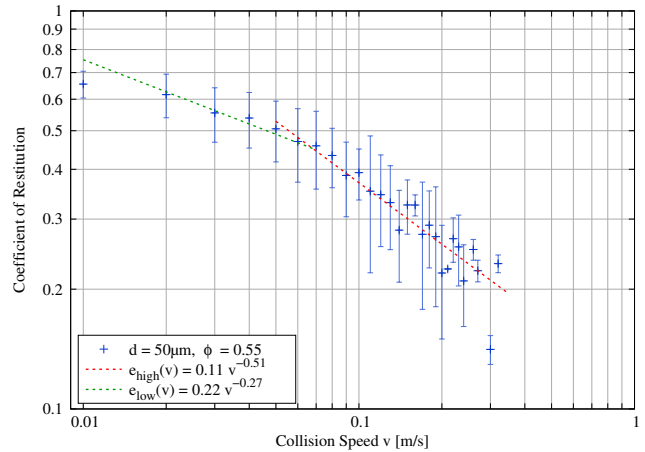}}
\caption{Dependency of the coefficient of restitution on the collision velocity obtained from molecular dynamics simulations. The collision partners are two identical $\approx 50\,\mathrm{\mu m}$ sized spherical dust aggregates having a mean porosity of $\phi \approx 0.55$.}
\label{fig:coefficient_of_restitution}
\end{figure}

The results of these collision sequences for the coefficient of restitution are shown in Figure \ref{fig:coefficient_of_restitution}. Similar to the laboratory experiments, one can distinguish different velocity regimes. The velocity dependency of the coefficient of restitution can be well fitted by power laws, $e(v) = a v^b$. We determined two fits $e_{\mathrm{low}}(v)$ for the low-velocity regime ranging from $1 \rm cm~s^{-1}$ to $7 \rm cm~s^{-1}$ and $e_{\mathrm{high}}(v)$ for velocities from $5 \rm cm~s^{-}$ to $33 \rm cm~s^{-1}$. We obtain $a_{\mathrm{low}} = 0.219$, $b_{\mathrm{low}} = -0.268$ and $a_{\mathrm{high}} = 0.109 $, $b_{\mathrm{high}} = -0.513 $. For the low-velocity regime, the exponent of the power law agrees very well with the one derived by \citet{ThNi} for the continuum theory, i.e. $b_{\mathrm{theory}} = -0.25$. The error bars show the influence of the orientation of the colliding aggregates on the outcome of the collision. For velocities higher than $33 \rm cm~s^{-1}$, we did not observe bouncing behavior anymore.

\begin{figure}[!h]
\resizebox{\hsize}{!}{\includegraphics{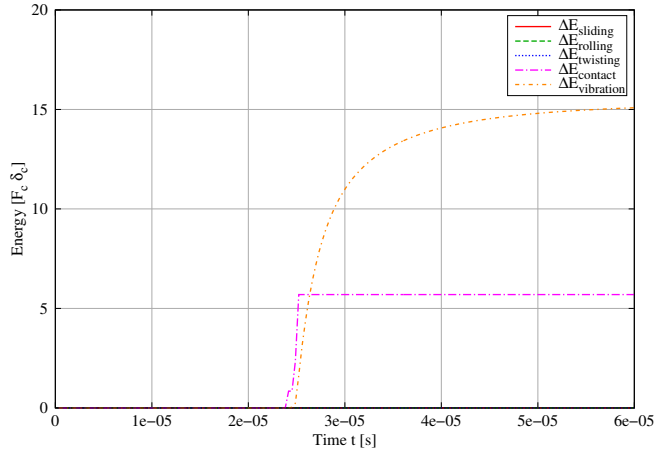}}
\caption{Energy dissipation during a bouncing collision in the low velocity regime with an impact velocity of 1 $\rm cm~s^{-1}$. The impact occurs after 23 $\mathrm{\mu s}$. The kinetic energy is dissipated by contact formation and breaking and vibrations of monomers.}
\label{fig:diss_energy_1}
\end{figure}

\begin{figure}[!h]
\resizebox{\hsize}{!}{\includegraphics{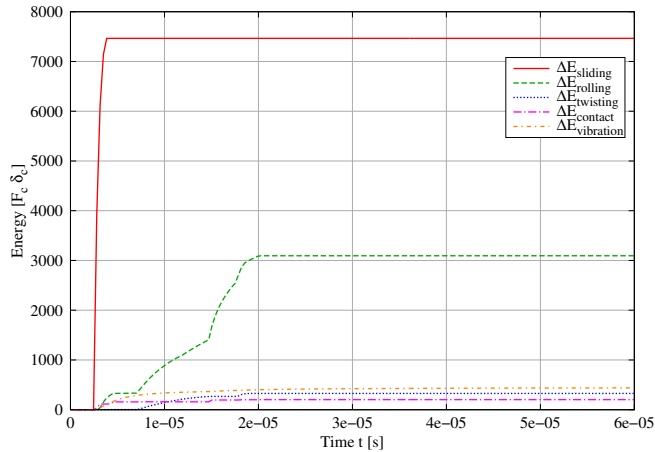}}
\caption{Energy dissipation of a bouncing collision in the high velocity regime with an impact velocity of 10 $\rm cm~s^{-1}$. The impact occurs after 2.3 $\mathrm{\mu m}$. Contrary to the low-velocity regime (Figure \ref{fig:diss_energy_1}), energy is mainly being dissipated by inelastic sliding and rolling.}
\label{fig:diss_energy_10}
\end{figure}

For low-velocity collisions, the kinetic impact energy is too low to restructure the aggregates and, thus, the energy is mainly dissipated by the formation and breaking of contacts between the monomers as shown in Figure \ref{fig:diss_energy_1}. At velocities above 2  $\rm cm~s^{-1}$, restructuring of the dust aggregate sets in. As we can see in Figure \ref{fig:diss_energy_10}, inelastic rolling and sliding become the most important dissipative channels. The dominance of the inelastic sliding results from the high compactness of the aggregates. Due to the large coordination number, monomers are tightly fixed by their neighbors, which limits the amount of inelastic rolling that may occur.

Given the vast differences in monomer-particle numbers between the experimental and the model dust aggregates, the agreement in the range of absolute values of the coefficient of restitution and the occurrence of different physical regimes is striking. Both approaches show a solid-state-like behavior for very low impact velocities and deviations from the expected power law with a slope of -1/4 for higher velocities. 

\section{ASTROPHYSICAL APPLICATIONS} \label{sec:applications}

\subsection{Protoplanetary Disks}
It has only recently been shown that cm-sized dust aggregates can grow in PPDs \citep{Zsometal2010}. Under certain conditions, even larger dust aggregates can be formed. \citet{Windmarketal2012} showed that a few indestructible, cm sized solid bodies can trigger the further growth of dust aggregates through a fragmentation and re-accretion cycle. Baroclinic vortices \citep{KlahrBoden2004} or streaming instabilities \citep{YoudinGoodman2005} can concentrate cm-sized or larger dust aggregates to high number densities. In all such cases, knowledge about the low- and intermediate-velocity collision behavior of dust aggregates is of utmost importance to correctly describe the fate of the dusty components. In highly mass-loaded regions of PPDs, such as the dust sub-disk or instability regions, the coefficient of restitution determines the reduction of the relative velocities among the dust aggregates. A low value of the coefficient of restitution eases the occurrence of gravitational instabilities. It is, thus, important to know the threshold velocity for the onset of fragmentation, so that the size evolution of the dust aggregates and, thus, the efficiency of the gravitational instability in PPDs can be correctly determined (see \citet{Johansenetal2006,Johansenetal2008,Johansenetal2012}).

\subsection{Debris Disks}
As relative velocities in virtually gas-free debris disk are typically larger than tens of meters per second, fragmentation dominates the outcomes in collisions between dust aggregates. Thus, one would expect a broad size distribution of dust aggregates from the monomer grains or the radiation-pressure blowout size (whatever is larger) to the largest occurring bodies. However, the recent discovery of ultra-cold debris disks with dust temperatures below the black-body equilibrium temperature has severely challenged this picture \citep{Eiroaetal2011}. Such low temperatures require dust materials with very low absorption in the far infrared and the absence of a source for particles smaller than the wavelength. The former can possibly be reached with icy constituents, the latter requires collision velocities below the fragmentation threshold.

We found in our investigation that the fragmentation limit is as low as $\sim 50\rm~cm~s^{-1}$ for large dust aggregates consisting of micrometer-sized $\rm SiO_2$ grains. This is very close to the $\sim 100\rm~cm~s^{-1}$ sticking threshold for the monomer particles \citep{PoppeBlumHenning2000}. Unfortunately, it has only recently been possible to produce micrometer-sized water-ice particles \citep{Gundlachetal2011} so that impact experiments with monomer particles and aggregates have not yet been performed. However, the experiments by \citet{Gundlachetal2011} showed that the surface force for the water-ice particles is about a factor of ten higher than for silica so that one can expect a similar increase for the threshold velocities for monomer sticking and aggregate fragmentation, i.e. $v_{\rm frag,ice} \approx 10~\rm m~s^{-1}$. Future experiments will have to show whether this is true or not.

\subsection{Planetary Rings}
In Saturn's main rings, the particle sizes have been estimated to range from $\sim 1$ cm to $\sim 10$ m \citep{Zebkeretal1985} and are believed to be at least covered by a regolith layer \citep{Pouletetal2002}, which makes them similar in their collision behavior to our dust aggregates. In the rings, the orbital shear leads to random velocities among the ring particles. This effect is counterbalanced by the energy loss due to the inelastic collisions among the ring particles, leading to a steady-state velocity distribution if and only if the coefficient of restitution decreases with increasing collision velocity. A basic description of this behavior is given in \citet{Goldreich1978}. The authors also found a direct correlation between the optical depth of the ring and the coefficient of restitution if the rings are in dynamical equilibrium. For reasonable optical depths for Saturn's main rings, the required coefficients of restitution are $e \stackrel{>}{\sim} 0.6$. Both conditions, the negative correlation between the coefficient of restitution and the collision velocity as well as the required values of the coefficient of restitution show that the physical composition of the ring particles must be such that the threshold velocity between elastic and plastic impacts is just slightly lower than the typical impact speeds ($\stackrel{<}{\sim} 1~\rm cm~s^{-1}$). Clearly, much more work has to be done before we can really understand the collision behavior of ring particles (e.g. use icy aggregates, much lower collision velocities, include rotation, etc.), but the dust-aggregate collision experiments presented in this article show how to proceed.

\section{SUMMARY}\label{kap:COCON}
We investigated the low-velocity collision behavior of 5-cm sized cylindrical dust agglomerates made by compression of micrometer-sized $\rm SiO_2$ particles. We measured the coefficient of restitution and the onset of fragmentation for agglomerates with volume filling factors of $\phi = 0.3$ and $\phi = 0.4$ and found that (1) at the very lowest velocities ($v_{\rm eff} \lesssim \rm 4~cm~s^{-1}$), the coefficient of restitution follows the solid state model by \citet{ThNi} within the measurement uncertainties as shown by the error bars in Figure \ref{fig:kofref}, that (2) the coefficient of restitution decreases steeply to $e = 0.2$ for a volume filling factor of 0.3 for velocities of $\rm 4~cm~s^{-1} \lesssim v_{\rm eff} \lesssim 6~cm~s^{-1}$ and to $e = 0.3$ for a volume filling factor of 0.4 for velocities of $\rm 4~cm~s^{-1} \lesssim v_{\rm eff} \lesssim 10~cm~s^{-1}$), that (3) the coefficient of restitution remains constant for higher velocities, until (4)  fragmentation dominates for $v_{\rm eff} \gtrsim 40 \pm 10 ~\rm cm~s^{-1}$. 

Our own numerical simulations, using the molecular-dynamics approach described in \citet{Seizingeretal2012}, yield a reasonable match to the experimental data over the entire bouncing regime. We discussed the consequences of our results concerning PPDs, cold debris disks, and planetary rings.

\bigskip
\bigskip

We thank Evelyn Liebert, Hans-Georg Pietscher and Heinrich Schmidt for their help in adjusting the laboratory drop tower to larger dust aggregates and for some test experiments. We thank Stephan Olliges and the Institut f\"ur Partikeltechnik at the TU-Braunschweig for performing the XRT mearsurements of our samples. This research was supported by the Deutsches Zentrum f\"ur Luft- und Raumfahrt under grant no. 50WM0936 and by the Deutsche Forschungsgemeinschaft (DFG) under grants BL 298/17 \& KL 650/16, and within the collaborative research group FOR~759 \textit{The formation of planets}. The simulations were performed on the bwGRiD cluster, which is funded by the Ministry for Education and Research of Germany and the Ministry for Science, Research and Arts of the state Baden-W\"urttemberg.

\bibliographystyle{aa}
\bibliography{ms}

\end{document}